\documentstyle[pre,aps]{revtex}
\begin{document}
\title
{Cooperative  Model of Bacterial Sensing}
\author{Yu Shi
%\cite{e}  
and Thomas Duke} 
\address{TCM Group,
 Cavendish Laboratory, Madingley Road, Cambridge CB3 0HE, UK}
\draft
\maketitle
\begin{abstract}
Bacterial chemotaxis is controlled by the signalling of a 
cluster of receptors. 
A  cooperative model is presented, in which coupling between neighbouring
receptor dimers enhances the sensitivity with which stimuli can be 
detected, without diminishing the range of chemoeffector concentration
over which chemotaxis can operate. Individual receptor dimers have two
stable conformational states: one active, one inactive. Noise 
gives rise to a distribution 
between these states, with the probability  influenced by
ligand binding, and also by the conformational states  of adjacent receptor 
dimers.
The  two-state model is solved,   based  on an
equivalence with the Ising model in a   randomly distributed 
magnetic field. The model has only two effective parameters,
 and unifies a number
 of experimental findings.
According to the value of the parameter comparing coupling and noise,
  the signal can be arbitrarily sensitive to 
changes in the fraction of  receptor
dimers to which ligand is bound.
The counteracting  
effect of a change of methylation level is mapped to an induced field 
in the Ising model. By returning the activity to the 
pre-stimulus level, this adapts the receptor cluster  to a new ambient
concentration of chemoeffector and ensures that a sensitive
response can be maintained over a wide range of concentrations.
\end{abstract}

\section*{I. INTRODUCTION}
One of the reasons why  we think  living things are special 
is that they show  awareness of the environment: they
respond sensitively  to stimuli and can adapt to 
 changes in the surroundings.   
Such biological complexity is displayed even by  bacteria, which,
in order to survive,
have to  be aware of their   precarious environment where  various conditions,
such as nutrient and toxin levels, acidity and
temperature,  may change rapidly \cite{parkinson}.
In order to discover universal principles, applicable at many levels of 
biological complexity, by investigating  a simple system, Adler revived  
 studies on bacterial chemotaxis which had been intensively 
investigated a century ago \cite{adler}. 
Recent 
 genetic engineering methods have made 
 it
  a paradigmatic system 
of  cellular signaling and adaptation.

A bacterium such as  {\em Escherichia coli} or {\em Salmonella typhimurium}
swims smoothly by rotating a bundle of
helical flagella  counterclockwise, but  tumbles 
 chaotically if the flagella rotate clockwise. When it moves towards 
a higher  concentration of attractant,  such as aspartate,  
 it  tumbles less frequently. When it moves towards a higher
 concentration of
repellent, it tumbles more frequently. So the bacterium performs a biased
random walk 
 towards
an attractant and  away from a repellent. 
This phenomenon is called chemotaxis \cite{stock1}. It 
is mediated by  receptors with 
extracytoplasmic sensing domains, connected by transmembrane helices to 
 signaling domains in the cytoplasm. The receptors, which are 
predominantly dimeric,   cluster  at one pole of the cell
 \cite{parkinson2}.  There are several types of
transmembrane receptors, which respond to different chemoeffectors
but
use the same signaling pathway \cite{stock1,falke,blair}, 
as shown in FIG. 1. 
Each receptor dimer 
is joined to two  CheA kinase molecules, via two 
 CheW proteins, forming a 
2:2:2 complex.
CheA is autophosphorylated at a rate that is
greatly enhanced by the receptor.
The phosphate is then passed from CheA to a regulator  protein,
CheY.  When  phosphorylated CheY diffuses to a
rotatory motor, 
the probability of clockwise rotation of the motor, and 
consequently  the tumbling frequency of the bacterium,
  increases.
The binding of a chemoeffector
 ligand to a receptor dimer
can trigger a 
transmembrane conformational change
which  regulates the autophosphorylation of CheA; 
  attractant binding decreases the rate, 
while   repellent binding increases it. In this way, an extra-cellular 
stimulus, i.e.  
ligand binding to   a receptor,  can modify the tumbling frequency.
It is  generally thought  that there exist two stable conformational
states of the receptor dimer:
 an ``active'' conformation  which corresponds to a very high rate
of CheA autophosphorylation; and
 an ``inactive'' conformation  which  corresponds to 
a lower rate.

An important feature of  chemotaxis is that the tumbling frequency
does largely depends  on
 recent changes of the concentration of chemoeffector.
This is achieved through an  adaptation process, which returns
the activity of the system  to  the pre-stimulus  level
after a period of time.
Adaptation is assured by a feedback loop which involves another regulator 
protein,
CheB. Like CheY, CheB also receives a phosphate group from CheA.
Phospho-CheB mediates a slow demethylation
of the receptor, countering the action of CheR, which promotes methylation.
Attractant binding also makes the receptor
a better substrate for CheR.
Since  methylation enhances the autophosphorylation rate of CheA, 
the change in the rate of phospho-CheY production
is gradually reversed by the feedback,
 and the the tumbling frequency ultimately 
returns to the pre-stimulus level. 

 Each subunit of a receptor dimer consists of two helices.
It has been proposed that the transmembrane signaling involves
a scissor-like   or pivoting 
 motion of the pair of  subunits \cite{milburn,cochran}, 
 or a piston-like  motion   involving
a conformation change within just one subunit 
\cite{milligan,pakula,stoddard}. The latter mechanism 
is favored by  recent 
distance-difference analyses of
the  aspartate receptor, which  reveal  that attractant
 binding induces  a displacement
of one helix, down towards the cytoplasm, 
while the other three helices  are not detectably perturbed 
\cite{chervitz,hughson}. 
  
The chemotactic response is extraordinarily  sensitive; as little as a single 
attractant molecule can trigger a detectable motor response 
\cite{block,berg,falke}. 
Recently, Bray, Levin and Morton-Firth suggested that this sensitivity
might be related to the clustering of receptors on the surface of the
bacterium. Without discussing any underlying mechanism,
they considered the possibility that the binding of a single 
molecule ligand  affects the activity of a number of receptors, so that
the response is augmented \cite {bray}.
In this article, we present a physical model of collective 
signalling in a cluster of receptors. We propose that 
the cluster responds as an entity, as a consequence of nearest-neighbor 
coupling between individual receptor dimers. In our model, the
 influence of one
dimer on another depends only on its activity, and not on whether it is
liganded. Noise causes each of the receptor dimers to fluctuate between
active and inactive states. When a given receptor dimer binds a 
molecule ligand,
 the probability of it being active is altered. Owing to the coupling,
the probabilities  of  activity of adjacent receptor dimers
 is also modified and this
effect propagates throughout the cluster. Thereby, the response to a 
stimulus is amplified. Moreover, ligand binding  is a rapid process,
while
which  of the receptor dimers are liganded is random, thus the above effect is
averaged out,  and the overall signal is a statistical average quantity 
as a function of the fraction of liganded receptor dimers. 
The model can be cast as an elementary neural network and reduced to
the Ising model. Thus the paradigmatic system of cellular
 signaling and adaptation is related 
to its counterpart in statistical mechanics. 
The model provides a simple, unifying framework to understand a 
large amount of experimental data. 
Sensitivity to very small changes of concentration, 
together with the ability to respond to gradients over a broad range of
concentrations, can naturally be attained. 
The model might be applicable to a variety of cellular signalling 
processes which demand a combination of low threshold of response
and wide dynamic range.

The  organisation of the
 rest of the article is  as follows. In Section II, we 
analyse various experimental
 results
and argue
for  the necessity of taking into account  inter-dimer coupling, as
well as  noise.
The neural network-like model is constructed
in Section III; it is solved in
Section IV, by  reducing it  to the Ising  model with a randomly distributed 
magnetic field.
Adaptation, and subsequent signaling of the adapted system, is
specifically discussed in Section V. Section VI contains 
discussions and   a summary. 

\section*{II. COUPLING AND NOISE}
\subsection*{A. 
COUPLING}
The existence of coupling among receptor 
dimers is indicated by a number of experimental results.
First,
cooperation among receptors in signaling and adaptation 
is hinted at by the fact that 
most of the chemotactic receptors cluster into a patch, located  at
 one pole of the cell \cite{parkinson2,falke}.
From the viewpoint of evolution,
we might 
 formulate a useful biological  principle:
{\em An attribute that exists 
 most probably confers advantages over possible
 alternatives, especially if the latter have some apparent merit.}
In the  present case, 
 if there were no cooperation among receptors,
a uniform distribution over the surface
would be optimal  in efficiency for  capturing
  molecules \cite{bergpurcell}.
Since, in practice, they are found to cluster together, there is most likely
 an advantage due to this feature.
Therefore,   coupling among the receptors might well
 play a role in 
signaling and adaptation. Coupling among receptor dimers can certainly
improve the
sensitivity. It can amplify
the signal generated by a stimulus, as has been anticipated by some authors. 

Second, 
it has been   found that  signaling  can occur through receptor 
dimers that have been genetically engineered so that one subunit
lacks a signaling domain
 \cite{milligan,gardina,tatsuno,stock}.
As mentioned above, a  conformational change of only one subunit
 has been observed
in the crystal structure 
\cite{chervitz,hughson}.
If the two subunits have essential differences and only
one of them
is involved
in the transmembrane signaling,
 then  inter-dimer coupling is inevitable to explain
the experimental results  on  truncated subunits (with $50\%$ probability
the truncation would have been 
  made on the signaling subunit, and no signaling would occur if 
the dimers act independently).
However, there is also the possibility  that the 
 binding of ligand to one subunit automatically suppresses 
 binding to 
the other subunit;
then 
the transmembrane signal can always be generated with either  subunit.
In this case,
inter-dimer coupling is not essential to explain 
the above experiments.
Coupling is, however,  necessary in order to reconcile 
the fact that  dimers with 
a truncated subunit are functional with
the favored mechanism of methylation, which requires interactions between two
subunits of the cytoplasmic domain \cite{falke}.

Third,  
it has been proposed,
 based on experiments, that  at least in certain cases, receptor
methylation is related to dimer-dimer interactions, i.e. methyltransferase
 bound to one dimer can methylate other dimers \cite{wu}\cite{falke}.
Another  support for coupling  is
 the remarkable mobility of the P2 domain of CheA, which
 provides the docking 
site for CheY and CheB.
 This mobility can serve to 
amplify the phosphorylation signal \cite{falke}.
Finally, 
 a recent {\it in vitro}  experiment by Liu {\it et al.}
showed  that kinase activation by a soluble signaling domain construct involves
the formation of a large complex, with about 14 receptor signaling domains
per CheA \cite{liu}.
This appears to be
a
strong 
 support for the coupling among receptor dimers {\it in vivo}. 

\subsection*{B.   NOISE}

A proper consideration of noise is important for several  reasons.
Thermal noise is 
certainly a significant issue since, for biological molecules, the energy 
barriers between distinct conformational states are generally comparable to 
 $kT$. Thus there is a high probability of stochastic 
transitions from one conformation to another. Indeed, experiments have 
indicated that there is considerable thermal motion in receptors. Disulphide 
trapping studies of the galactose/glucose binding protein have revealed 
spontaneous, large amplitude thermal fluctuations of the protein backbone 
structure [5].

Moreover, noise can also provide benefits. In the absence of noise, 
nearest-neighbour coupling amongst receptor dimers would cause the activity
to spread across the
whole array and would inevitably make the response to different stimuli
indistinguishable . If noise is taken into account, individual receptors
flicker between active and inactive states
and the overall signal is a statistical average, which 
naturally varies for different numbers of liganded receptor dimers.

With the consideration of  noise, it is worth emphasising  that the
signaling process should be understood within the context of 
dynamic equilibrium:
When  the concentration of chemoeffector is stable,
the conformational state and the level of methylation of each 
receptor dimer fluctuates
microscopically, but the mean activity of the system
remains unchanged. This equilibrium is shifted when the concentration of 
chemoeffector is changed. 

\section*{III. THE MODEL}
We  study the total signal  
 of the  cluster of receptors  
 as a quasi-equilibrium property corresponding  to a certain 
concentration
of chemoeffector and 
a certain  level of methylation. 
This approach is justified by the   
wide separation of time scales in this system:
Ligand binding and protein conformational transitions occur within
milliseconds. Changes in protein phosphorylation occur on a time scale 
of $\sim0.1s$. The much slower 
 adaptation process, associated with modulations
of  methylation level, is 
  on a scale of  minutes  \cite{stock1,springer}.  

The quasi-equilibrium state of a dynamics is  determined by the minimum of
a noisy  ``energy function'' (a Lyapunov function).
This ``energy''  is not necessarily  the actual
physical energy,  since it may describe  an effective dynamics
that
``renormalizes'' the underlying chain of physical processes. 
Similarly, the noise may  not be due only to
 the temperature, but in the present case it mostly is. 
Such a description, 
which reduces  degrees of freedom,
 is  especially effective when the underlying physical processes are 
 complicated, or unclear in detail.
A typical example is  Hopfield's neural network model
 \cite{hopfield,amit}. Here we  adopt this approach for chemotactic  signaling,
 but  with a different 
interpretation and 
with the emphasis
 placed on the determination of the equilibrium activity
as a function of  the 
 external stimuli.

Consider a lattice of receptors,
whose basic
unit is the 
receptor dimer, or equivalently, the whole receptor-CheW-CheA dimer
 complex. 
Initially, we consider a system that has adapted to zero 
concentration of the chemoeffector, and investigate
the response when the concentration changes so that
a fraction $c$ of receptor dimers are  bound to chemoeffector molecule ligands.
The state of a receptor is a function of the effect of 
ligand binding and the states of 
the neighboring receptor dimers.
Characterizing the state of  receptor dimer $i$ by a variable $V_{i}$ 
(according
to  recent observation, 
it is  the vertical  position of 
one of four helices, but we are open to possible 
re-interpretation according to experimental findings),
 and the
effect of the ligand binding to receptor dimer $i$ by $H_{i}$,
 most generally we have 
\begin{equation}
V_{i}\,=\,V_{i}(\{V_{j\neq i}\},\{H_{j}\}),
\end{equation}
where  $\{H_{j}\}$ denotes the set of $H_{j}$ for 
 $j\,=\,1,2,\cdots$, $\{V_{j\neq i}\}$ denotes the set of all $V_{j}$ for
 $j\,\neq\,i$.
 The natural assumption is that $V_{i}$ is 
affected only  by  $H_{i}$ and the states of the nearest neighbours. 
Furthermore, for the 
two-state model, in which $V_{i}$ has only two possible values, 
$V^{0}$  or $V^{1}$, the  McCulloch-Pitts threshold
 model \cite{mp,amit}
 is a natural assumption. 
Thereby 
\begin{equation}
V_{i}\,=\,\psi(\sum_{j}T_{ij}V_{j}+H_{i}-U_{i}),
 \mbox{with}\,
\psi(x) \,=\,\left\{  \begin{array}{ll}
V^{1} & \mbox{if $x>0$}\\
V^{0} & \mbox{if $x\leq 0$}
\end{array}\right. ,
\end{equation}
where $U_{i}$ is a threshold
value, and $T_{ij}$ describes the coupling  among receptor dimers, which  is 
assumed to be nonzero only for nearest neighbours.
We adopt the convention that $V^{0}$ is the active conformation and
$V^{1}$ is the inactive one. Then $H_{i}\,>\,0$ for attractant binding,
which tends to inactivate receptors and 
depress the autophosphorylation rate of CheA, thereby decreasing
the frequency of tumbling. Conversely, $H_{i}\,<\,0$ for 
repellent binding. 

   It is well known that if 
$T_{ij}\,=\,T_{ji}$ and $T_{ii}\,=\,0$,  clearly
valid in the present situation, the dynamics is determined by a 
Lyapunov function (or Hamiltonian) \cite{amit},
\begin{equation}
{\cal H}\,=\,-\sum_{<ij>}T_{ij}V_{i}V_{j}-\sum_{i}H_{i}V_{i}
+\sum_{i}U_{i}V_{i},\label{hamiltonian}
\end{equation}
where $<ij>$ represents pairs of nearest neighbours.
Taking into account the noise, which induces  a state  distribution
 which 
is nearly 
a Boltzmann distribution \cite{amit}, the problem reduces to
the  statistical mechanics 
of a system with the above   Hamiltonian.

In the simplest interpretation, the noise is purely thermal, 
$\beta=1/kT$, and Eq. (\ref{hamiltonian}) may be identified as
the effective physical energy. According to recent observation,
$V^0$ and $V^1$ are 
the two  stable positions  of one of the four helices. 
Therefore $H_{i}$ and $T_{ij}V_{j}$
are forces due to ligand binding and coupling, respectively. 
The ``Zeeman energy'' dependent on ligand binding is due to the free energy 
exchange with bound ligand. Similarly, the coupling energy is due to free
energy exchange with the cytoplasm or membrane,
 which mediates  the effective coupling. 
   
Eq. (\ref{hamiltonian})
may be  transformed  to the ``spin'' representation by writing
$S_{i}\,=\,2(V_{i}-V^{0})/\Delta V-1$, where $\Delta V\,=\,(V^{1}-V^{0})$.
Then 
\begin{equation}
{\cal H}
\,=\,-\sum_{<ij>}J_{ij}S_{i}S_{j}-\sum_{i}B_{i}S_{i}+{\cal H}_{1}+E_0,
\label{hamiltonian3}
\end{equation}
where  $J_{ij}\,=\,T_{ij}\Delta V^2/4$, $B_{i}\,=H_{i}\Delta V/2$, and 
$E_{0}$ is
a constant given a distribution of $\{B_{i}\}$. 
${\cal H}_{1}$
is a  ``Zeeman energy'' due to an effective ``magnetic field''
independent of    $\{B_{i}\}$, which  determines  the  equilibrium 
configurations in the absence of  $\{B_{i}\}$, i.e. without ligand 
binding. Without loss of 
generality, we  
set ${\cal H}_{1}\,=\,0$. Thus in the absence of
$\{B_{i}\}$,  and  if the noise is sufficiently 
high, $S_{i}$ is equally distributed between $1$ and $-1$, 
and  the ``magnetization'' is zero. In other words, it is assumed that
there is no energy difference between  the active $(S_i = -1)$ and
inactive $(S_i = 1)$ conformations for an isolated, 
unliganded receptor dimer $i$.  The physics does not change if this difference
is set to be nonzero.
Ligand binding shifts the  energy   difference  to $2B_i$. 

We have now reduced the model to an Ising model. The activity of the 
array of receptors corresponds to the average magnetization of a lattice
of spins, and ligand binding  of a receptor dimer corresponds to a local
magnetic field at a lattice site:
$B_{i}\,=\,B$ if  receptor dimer $i$ binds a chemoeffector 
ligand, otherwise  $B_{i}\,=\,0$.
If the fraction of liganded receptor dimers 
is $c$, then the value of $B_{i}$  is randomly distributed between $B$ and
$0$ with  probability 
\begin{equation}
p(B_{i})\,=\,c\delta(B_{i}-B)+(1-c)\delta(B_{i}).\label{dis}
\end{equation}
 This Ising  model
 in a field bimodally distributed
 between $0$ and $B$ is simpler 
than    the so-called 
``random-field Ising model''   \cite{imry,schneider,aharony,young},
 in which the possible
values of the field  are  symmetric with respect to zero,
  and   nontrivial results arise due to  the fluctuation of the
fields. In our case, the average of the 
 field is nonzero, so there is  a long-range
order simply as the result of the explicit  symmetry breaking. 
For Eq. (\ref{hamiltonian3}), 
$\overline{B}_{i}\,=\,cB$, where the overbar denotes the average over 
disordered
configurations. The fluctuation of the  random distribution
 is $\Delta B_{i}\,=\, \sqrt{c(1-c)}B$.
Consider  the formation of a domain of size $L$
in the ferromagnetic ground state. According to the central limit theorem,
the average Zeeman energy is $\sim L^{d}cB$, much larger than
its fluctuation, which is
$\sim L^{d/2}\sqrt{c(1-c)}B$. Therefore the energy gain is always positive,
and the fluctuation of the field   cannot destroy   long-range order.

\section*{IV. SOLUTIONS OF THE  MODEL}
The two-state model, which has been reduced to the Ising model
in  a randomly distributed field, as  described by Eqs. (\ref{hamiltonian3})
and (\ref{dis}),   can be solved by the  mean-field method.
One may obtain the result simply by considering that the
 average magnetization,
  $m\,=\,\overline{<s_{i}>}$,
where $<\cdots>$ denotes the thermodynamic average, 
 is determined by the
local field $B'_{i}\,=\,\sum_{j}J_{ij}m+B_{i}$ with the random
distribution.  Alternatively one may first obtain the free energy using 
the replica method, then calculate 
the average magnetization \cite{schneider,aharony}. 
It is found that $m$ is the root of the equation
\begin{eqnarray}
m&=&\overline{tanh(\beta \nu Jm+\beta B_{i})}\label{ave}\\
&=&\frac{2c}{1+\exp[-2(\beta\nu Jm+\beta B)]}
+\frac{2(1-c)}{1+\exp (-2\beta\nu Jm)}-1,\label{mag}
\end{eqnarray}
while the the noisy Lyapunov function is
\begin{equation}
F\,=\,\frac{1}{2}\nu Jm^{2}
-\frac{1}{\beta} \{c\ln [2\cosh(\beta\nu Jm+\beta B))]
+(1-c)\ln [2\cosh(\beta\nu Jm)]\}+E_{0}.
\end{equation}
Here,  $J_{ij}$  has been assumed to have  a single value 
$J$ 
for nearest neighbors, $\nu$ is the number of nearest neighbors,
 and $\beta$ is a characterization of the noise. 

The relation between the chemoeffector concentration and
the activity of the 
system  is now reduced to the  $m$ versus $c$ relation,
 determined by Eq. (\ref{mag}), since the 
 activity of the system, here defined as the fraction of receptor dimers
in the active state  is
$A\,=\,(1-m)/2$, and the pure response to the stimulus, i.e.
the change  of the activity, 
 is $\Delta A\,=\, m/2$. 
Although   Eq. (\ref{mag}) may possibly have more than one 
solution, the one corresponding to the 
 lowest  $F$ is what we need.
 Approximate analytical solutions may be found in limiting cases,
\begin{equation}
m\,\approx\,\left\{ \begin{array}{ll}
 \frac{\beta c B}{1-\beta\nu J} &\mbox{if}\,  \beta\rightarrow 0\\
 1-2(1-c)\exp{(-2\beta\nu J)}-2c\exp{[-2(\beta\nu J+\beta B)]}&\mbox{if}\,  
\beta\rightarrow \infty \,\mbox{and} \,B > 0\\
 -1+2(1-c)\exp{(-2\beta\nu J)}+2c\exp{-2[\beta\nu J+\beta|B|)}& \mbox{if} \, 
\beta\rightarrow \infty \,\mbox{and} \,B <  0
\end{array}\right.\label{limit}
\end{equation}
In general, the solution  can only be obtained numerically.
It can be seen that there are actually only two effective
 parameters in this model; 
one is $\alpha=\,\beta\nu J$, the other is $\gamma\,=\, \beta B$.
Owing to symmetry, it suffices to give  results for $B\,>\,0$.
Solutions for typical values of parameters are shown in 
 FIG. 2.
First we choose 
 $\alpha\,=\,0.1,\,0.8,\,
1.2$; then  for each $\alpha$, the dependence of $m$ on $c$ is 
 determined for  $\gamma\,=\,0.01,\,0.01,\,0.1,\,1,\,10,\,100$.
% also for $\gamma\,=\,1000$ when $\alpha=0.1$.
 Note that for $c\,=\,0$, i.e., the Ising model without
a magnetic field, $\alpha\,=\,1$ is the critical value dividing
the ``paramagnetic'' phase, where $m(c=0)\,=\,0$,
 and the ``ferromagnetic'' phase, where there is a
 ``spontaneous magnetization'',  $m(c=0)\,\neq\,0$.

 The quantitative measure of 
sensitivity, denoted by $S$, is half of the slope at $c = 0$:
\begin{eqnarray} 
S&=&
\frac{\partial (m/2)}{\partial c}|_{c=0} \nonumber
\\
&=&
\frac{\frac{1}{1+\exp [-2(\beta\nu Jm_0+\beta B)]}-
\frac{1}{1+\exp(-2\beta\nu Jm_0)}}{1-\frac{4\beta\nu J\exp(-2\beta\nu Jm_0)}
{[1+\exp(-2\beta\nu J m_0)]^2}},
\end{eqnarray}
where $m_0=m(c=0)$.
It is clear  that $S$ can be made {\em arbitrarily} large
by choosing appropriate value of
$\beta\nu J$ so that the denominator in the above expression
is arbitrarily close to $0$. For $m_0=0$,
\begin{equation} 
S\,=\,
\frac{\frac{1}{1+\exp (-2\beta B)}-\frac{1}{2}}{1-\beta\nu J},
\end{equation}
which 
 is directly tuned by the difference between $\beta\nu J$ and 
$1$, which is the critical value of phase transition for $c=0$.
The case with $m_0 \neq 0$ is less favored since the range of
possible $m$ for different $c$ could be diminished, making it 
more difficult to distinguish between different stimuli.
Moreover, the sign of $m_0$ would be  
determined by that of the previous  $B$, conflicting the fact that 
the pre-stimulus level is fixed.  

For a given $\alpha$, $S$  also increases with $\gamma\,=\,\beta B$, 
but with an upper bound. The fact that 
$\partial m/\partial \gamma\,\rightarrow\,0$
when $\gamma\,\rightarrow\,\infty$ indicates that,
if ligand binding has a strong enough effect, the response 
is independent of the exact value of $\gamma$. This
provides a sort of stability for the effect of ligand binding.

Thus good sensitivity requires fine tuning of the coupling: the greater
the sensitivity demanded by the bacterium, the more accurately
$\alpha\,=\,\beta\nu J$ has to be controlled.
But 
$\gamma = \beta B$ may vary widely without considerably affecting the response.
This is reasonable, since the temperature range suitable
for bacterial survival is
rather restricted and, for a given 
bacterium,  $\nu J$ is a structural property, which could be
optimized during evolution. On the other hand,
the effect of ligand binding,  $B$, 
depends on the external stimulus, which
may vary considerably.  
  
As an exercise,  our model may be applied 
to the puzzling situation in which both
attractants and repellents are present \cite{adler}. 
In this case,
\begin{equation}
p(B_{i})\,=\, c_{r}\delta(B_{i}-B_{r})+c_{a}\delta(B_{i}-B_{a})
+(1-c_{r}-c_{a})\delta
(B_{i}),
\end{equation}
where $c_{r}$ and $c_{a}$ are the concentrations of the repellents and 
attractants, respectively, and $B_{r}$
and $B_{a}$ are respectively the measures of the attractant and repellent
 binding.
Obviously the activity is dependent on both $c_{r}$ and $c_{a}$.

\section*{V. ADAPTATION}
Now we incorporate into this model the delayed adaptation due 
 to the change of  methylation level.
This  may be achieved through an induced  ``field'' with an opposite
sign to that associated with ligand binding,
so that the 
``magnetization'' returns  toward the pre-stimulus level.
 This
assumption for the additivity of the effect of ligand binding and 
that of the
change of methylation level is supported by the finding that in a receptor
there is
a region which gathers, integrates, and interprets the multiple inputs 
transferred  by the transmembrane signaling domain and the methylated 
side chains, then transmits an output signal to 
the kinase regulation machinery \cite{falke}.

Two points should  be made clear. First,
the time scale of the change of the level of methylation of the whole 
system is much longer than the microscopic time scale, so the ``magnetization''
can still be  obtained as the  equilibrium property of the noisy 
Lyapunov function, which quasi-statically 
changes with the  level of methylation.
Second,  since the  ligand  binding   occurs on a 
time scale  much shorter than the time needed for  adaptation
to be completely achieved,
 we cannot simply change  the value of $B$, but must introduce
another ``field''.
We denote  this ``induced field'' 
by  $\{M_{i}\}$,  with the distribution
\begin{equation}
p(M_{i})\,=\,c_{m}\delta(M_{i}-M)+(1-c_{m})\delta(M_{i}),
\end{equation}
where $c_{m}$ is the fraction  of the receptor dimers 
whose original
level of methylation is modified. The sign of $M$ is
 opposite to that of $B$.
Thereby the net field is $D_{i}\,=\,M_{i}+B_{i}$ with the distribution
\begin{eqnarray}
p(D_{i})\,&=&\, cc_{m}\delta(D_{i}-B-M)+c(1-c_{m})\delta(D_{i}-B) \nonumber\\
  & &+(1-c)c_{m}\delta(D_{i}-M)+(1-c)(1-c_{m})\delta(D_{i}). \label{dd}
\end{eqnarray}
 The equilibrium state can be obtained by replacing $B_{i}$ in 
Eq. (\ref{hamiltonian3}) with $D_{i}$. Adaptation is taking place if
$c_{m}$  and/or $M$ vary  slowly with time.
This gives rise to a time-dependent ``magnetization'',
which  may return to zeo. 
To get a quantitative impression,
 by  adopting the high 
noise  limit
$\beta\,\rightarrow\,0$, 
it may be estimated that when
$c_{m}B+cM=0$,  adaptation is completed; the ``magnetization'' returns 
to zero. Here, we shall simply assume that a molecular mechanism
exists which ensures that the state of zero ``magnetization'' is an
attractor of the dynamics, so that adaptation is exact.
A more precise study of the adaptation process will be reported in 
the future.  

Once the system has adapted,
suppose that the concentration subsequently  changes from $c$ to $c+\delta c$.
One can obtain the new activity  by substituting in Eq. (\ref{dd}) 
$c+\delta c$ for $c$, and  the  values 
of $c_{m}$ and $M$   at which the 
  adaptation was completed. 
In general,
what is most important  is the  change of fraction  of liganded 
 receptor dimers
since  the last   adaptation.  
Under  high temperature approximation,
the  result is  
   Eq. (9) with $c$ replaced by $\delta c$.
Moreover, it can be seen that if  $\delta c$ is negative, 
i.e  if chemoeffector  is removed, the 
effect is similar to the addition of a chemoeffector  whose  ``field''
has opposite sign.
Therefore the removal of attractant is equivalent to repellent
binding, and vice versa. This has
indeed been   observed in experiments
 \cite{springer}.

\section*{VI. SUMMARY AND DISCUSSIONS}

In this article, we  analyse relevant experimental 
results and draw the conclusion that both inter-dimer coupling and noise 
are crucial in the mechanism of chemotactic signaling and adaptation.
The ratio between their measures, $\alpha\,=\,\beta\nu J$ is one of the two 
effective 
parameters in the cooperative model we construct. A second parameter 
is the ratio between the measure of the effect of 
ligand binding  and that of 
the noise, $\gamma\,=\, \beta B$. 
The essential features due to the balance of  coupling and noise
 are  well captured by the paradigmatic model of
statistical mechanics, the Ising model. We made an attempt to  map the
underlying mechanism  of  collective effects  in
 chemotactic signaling to the Ising model in a
randomly distributed field, with the distribution  reflecting  the ligand 
occupancy.
To complete the mapping, we adopted  the  basis of Hopfield's neural network 
model. The great  difference between time scales of the various
chemical and mechanical processes
 makes it feasible to obtain the signaling level as 
a quasi-equilibrium property of a noisy  Lyapunov function.
This Lyapunov function describes the dynamics ``renormalizing'' underlying
complexity.

Our model provides the following picture. An individual 
receptor dimer has two stable 
conformational states, 
an active one that corresponds to a high 
rate of CheA autophosphorylation, and an inactive one that 
corresponds to a low rate. Noise gives rise to a distribution between
these states and the partition is influenced both by ligand binding
and by the conformational states of the neighboring receptor dimers.
In the simplest interpretation, the noise is purely thermal, $\beta = 1/kT$,
$2B$ corresponds to the  shift of the energy difference between  
active and inactive states induced by ligand binding, and
$J$ measures the effective coupling energy 
between neighboring  receptor dimers. 
The activity of the receptor cluster is a statistical average quantity.
 A  change in the fraction of
liganded receptor dimers causes the total activity to change from 
the pre-stimulus level. But the level of 
methylation also changes, on a slower time scale. 
This causes an effect  opposite to that induced by ligand binding.
Consequently, the total activity ultimately returns to the pre-stimulus level.

The  coupling between receptor dimers naturally provides
the sensitivity to small stimuli observed in  experiments.  Additionally,
the noise makes the response to different values of 
concentration changes distinctive. Sensitivity to small changes in the
environment requires fine tuning of the parameter $\alpha$,
but $\gamma$ may vary without considerably affecting the response.
The equivalence between the removal of attractant  and the
addition of repellent, or vice versa, has a natural explanation.

Among  problems for further investigation are the effects of finiteness
of the number of receptor dimers, potential
randomness in the coupling, and features that might be
lost in the mean field solution.  Correlation between $B_{i}$, 
or $M_{i}$, at different sites $i$ is also a possibility and
might have useful consequences. 
The finite-size effect and the ``random field'' due to
the change of methylation level may destroy the
``spontaneous magnetisation''
 that exists   for $\alpha\,>\,1$,
thus relaxing the constraint on the precision to which $\alpha$ must be 
specified to give high sensitivity. The mean-field solution 
 is least accurate when $c \rightarrow
\frac{1}{2}$,
since the fluctuation of the field is $\sqrt{c(1-c)B}$, which increases to 
the 
greatest as $c \rightarrow
\frac{1}{2}$.  Thus maybe  the sensitivity is
 lower at moderate values of the occupancy $c$, than at the
extremes $c \rightarrow 0$ and $c \rightarrow 1$. However, this is
not necessarily a limitation. 
The fractional occupancy $c$ is related to the ambient concentration of 
ligand $[L]$ by
\begin{equation}
c = \frac{[L]}{[L]+K_d},
\end{equation}
where $K_d$ is the dissociation constant. Thus
\begin{equation}
\delta c\,=\,\frac{[L] K_d}{([L]+K_d)^2}\frac{\delta [L]}{[L]}\,
=\,c(1-c)\frac{\delta [L]}{[L]}.
\end{equation}
Given that the bacterium probably needs to detect a  relative change in
concentration, $\frac{\delta [L]}{[L]}$, we see that the greatest
sensitivity to a change in occupancy is demanded when  $c \rightarrow 0$ or
 $c \rightarrow 1$, and the least when $c \rightarrow  \frac{1}{2}$.

It is well known that the 
the two-state threshold neural network model
is equivalent to a model with continuous variables
 in the high gain limit \cite{hopfield},
with the Lyapunov function
\begin{equation}
{\cal H}\,=\,-\sum_{<ij>} T_{ij}V_{i}V_{j}+\sum_{i}\frac{1}{R_{i}}\int_{0}
^{V_{i}}g_{i}^{-1}(V)dV -\sum_{i}{H}_{i}V_{i}, \label{neu}
\end{equation}
with  $u_{i}\,=\,g_{i}^{-1}(V_{i})$  determined  by
\begin{equation}
C\frac{du_{i}}{dt}\,=\,\sum_{<ij>}T_{ij}V_{j}-\frac{u_{i}}{R_{i}}
+{H}_{i}, \label{tran}
\end{equation}
where  $u_{i}$ is interpreted as 
the soma potential, while $C$ is the input capacitance of the cell
membrane. When Eqs. (\ref{neu}) and (\ref{tran}) are adopted for
the network of 
 chemoreceptor dimers, $V_{i}$ is a variable characterizing
the stable 
conformation, i.e. the (vertical) position of the mobile helix of the
receptor dimer,  and $u_{i}$  is the instantaneous position.
Thus Eq. (\ref{tran}) could be  the equation of motion 
 describing  the transient  process
of the movement of the mobile helix, in response to a force ${H}_{i}$
generated by ligand binding, as well as forces due to couplings with 
the neighboring receptor dimers.  Of course, whether modification of
(\ref{tran}) is necessary depends on future experimental results.

According to this interpretation, by measuring the force generated by
ligand binding, ${H}$, and the displacement of the mobile
(signalling) helix
$\Delta V = (V^{1}-V^{0})$, one may obtain the parameter 
$B\,=\,{H}\cdot \Delta V/2$. Note that ${H}\cdot \Delta V$ is the work 
done by the force $H$, consistent with the identification of
$2B$ with the shift in  energy difference caused by ligand binding.
Similarly, $4J/\Delta V = T_{ij} \Delta V$
is the force generated by the conformational change of one of the
nearest neighbours.
To make a rough estimation, we take typical values $\alpha=0.5$, 
$\gamma=5$ and $1/\beta\approx 4pN\cdot nm$ (assuming that the noise
is purely thermal). Then $B\approx 20 pN\cdot nm$,
 $\nu J\approx 2 pN\cdot nm$. 
The measured displacement is $0.16nm$ \cite{chervitz}. It is found
that the force resulting from ligand binding is about 
$250pN$ and the force due to coupling between a pair
of nearest neighbours is about $10pN$. 
These orders of magnitude are quite reasonable. 

Since the  continuum model can be realized in electric circuits,
more insights might be provided from the 
viewpoint of system control, where negative feedback has been well
studied. On the other hand,
the analogy with the neural network model is possibly more than 
a mathematical
one. From the viewpoint of evolution,
 there are   common features between bacterial 
sensing and sensing of higher animals.
Perhaps a primitive or ancestral neural network works in chemotaxis.
 Adler writes: ``The basic elements that make behaviour
 possible in higher organisms are also present in a single
 bacterial cell; they are sensory receptors,
a system that transmits and processes sensory information and effectors to
 produce
movement. Whether the mechanisms of any of these elements in bacterium 
are similar to those in more complex organisms remains to be established'' 
\cite{adler}.   Margulis thinks: `` Thought and behaviour in people
are rendered far less mysterious when we realize that choice and sensitivity 
are already exquisitely developed in the microbial cells that became our 
ancestors'' \cite{margulis}.
We hope our approach is a small step  in addressing these issues.

\acknowledgements
We are very grateful to D. Bray for valuable  discussions and  comments, and  
for critically  reading  the manuscript.
Y.S.  also thanks G. Fath and P. Littlewood for discussions. 

\begin{figure}
\caption{A schematic illustration of the  chemotactic signaling pathway.}
\end{figure}

\begin{figure}
\caption{The solution of the two-state model:
 ``magnetization'' $m$ 
as a function of  the fraction $c$ of
liganded receptor dimers. 
Here we assume there was no ligand bound previously. 
The three figures are for three
typical values of the parameter $\alpha=\beta\nu J$:
(a) $\alpha\,=\,0.1$; (b)  $\alpha\,=\,0.8$;
(c)  $\alpha\,=\,1.2$.
In each figure, different plots are for different values of the  parameter
$\gamma\,=\,\beta B$: 
$\Diamond:\gamma=0.01;\,+:\gamma=0.1;\, \Box:\gamma=1; \, \times: \gamma=10$;
 $\bigtriangleup:\gamma=100$.
Note that the critical pont, which separates  ``ferromagnetic'' 
and ``paramagnetic'' phases, is  $\alpha\,=\,1$.}
\end{figure}


\begin{references} 
%\bibitem[*]{e} Electronic address: ys219@phy.cam.ac.uk       
\bibitem{parkinson} J.S. Parkinson and E.C.  Kofoid,
 {\em Annu. Rev. Genet.} {\bf 26},
71 (1992).
 \bibitem{adler} J. Adler,
  {\em Sci. American}, {\bf 234}(4), 40 (1976).
\bibitem{stock1} J. Stock and M. Surette, in {\em Escherichia coli and
 Salmonella typhimurium: Cellular and Molecular Biology}, ed. F.C. Neidhardt,
(ASM, Washington, 1996).
   \bibitem{parkinson2} J.S. Parkinson, J. S. and D.F. Blair,
 {\em Science} {\bf 259}, 1701 (1993); M.R.K. Alley, 
J.R. Maddock,  and L. Shapiro,
 {\it ibid.} 1754
 (1993); J.R. Maddock and L.  Shapiro,
 {\it ibid.} 1717 (1993).
\bibitem{falke} J.J. Falke {\it et al.}, 
{\em Annu. Rev. Cell Dev. Biol.}
{\bf 13}, 457 (1997).
\bibitem{blair} D.F. Blair, {\em Annu. Rev. Microbiol.} {\bf 49} 489 (1995).
\bibitem{milburn} M.V. Milbourn {\it et al.}, {\em Science} {\bf 254}, 1342
(1991).
\bibitem{cochran} A.G. Cochran and P.S.  Kim, 
{\em Science} {\bf 271}, 1113 (1996).
\bibitem{milligan}  D.L. Milligan and D.E.  Koshland Jr.,
 {\em Science} {\bf 254},
1651 (1991).
 \bibitem{pakula} A. Pakula and M.  Simon, 
 {\em Nature},
{\bf 355} 496 (1992).
\bibitem{stoddard} B.L. Stoddard, J.D. Bui, and D.E. Koshland Jr.,
{\em Biochemistry} {\bf 31}, 11978 (1992).
  \bibitem{chervitz} S. Chervitz and J.J.  Falke, 
{\em
Proc. Natl. Acad. Sci. USA} {\bf 93}, 2545 (1996).
\bibitem{hughson} A.G. Hughson and G.L. Hazelbauer, {\em Proc. Natl. Acad.
Sci. USA} {\bf 93}, 11546 (1996).
\bibitem{block} S.M. Block, J.E. Segall, and H.C. Berg, 
{\em J. Bacterial} {\bf 154}, 312 (1983).
\bibitem{berg} J.E. Segall, S.M. Block, and H.C. Berg, 
{\em Proc. Natl. Acad. Sci. USA} {\bf 83}, 8987 (1986).
\bibitem{bray}   D. Bray, M.D. Levin, and C.J. Morton-Firth,
{\em Nature},  {\bf 393}, 85 (1998). 
\bibitem{bergpurcell} H.C. Berg and E.M. Purcell, {\em Biophys. J.} {\bf 20},
193  (1977). 
  \bibitem{gardina} P.J. Gardina and M.D. Manson,
  {\em Science} {\bf 274}, 425 (1996).
 \bibitem{tatsuno} I. Tatsuno, M. Homma, K. Oosawa, and I. Kawagishi,
 {\em Science} {\bf 274}, 423
(1996).
 \bibitem{stock} J. Stock,
  {\em Science} 
{\bf 274}, 370 (1996).
\bibitem{wu} J.R. Wu, J.Y. Li, G.Y. Li, D.G. Long, and R.M. Weis,
{\em Biochemistry} {\bf 35}, 4984 (1996).
\bibitem{springer} M.S. Springer, M.F.  Goy,  and J. Adler, 
   {\em Nature} {\bf 280}, 279 (1979).
\bibitem{liu}  Y. Liu, M. Levit, R. Lurz, M.G. Surrette,  and J.  Stock,
{\em The EMBO Journal} {\bf 16},
7231 (1997). 
\bibitem{hopfield} J.J. Hopfield,
 {\em Proc. Natl. Acad. Sci. USA} {\bf 81}, 3088 (1984);
 {\em Phys. Today}, {\bf 47}(2), 40(1994).
 \bibitem{amit} D.J. Amit, 
{\em Modeling Brain Function} (Cambridge University Press, Cambridge, 1989).
\bibitem{mp} W.S. McCulloch and W.A. Pitts,
 {\em Bull. Math. Biophys.} {\bf 5},  115 (1943).
\bibitem{imry} Y. Imry and S.-K. Ma, 
{\em Phys. Rev. Lett.} {\bf 35}, 1399 
(1975).
 \bibitem{schneider} T. Schneider  and E. Pytte, 
 {\em Phys. Rev. B},
 {\bf 15}, 1519 (1977). 
\bibitem{aharony} A. Aharony,
 {\em Phys. Rev. B} {\bf 18}, 3318 (1978).
 \bibitem{young}  D.P. Belanger and A.P. Young, 
{\em J. Mag. Mag. Materials} {\bf 100}, 272 (1991).
\bibitem{margulis} L. Margulis,   
in {\em The Third Culture}, ed. J. Brockman,
(Simon \& Schuster, NewYork, 1995).
\end{references}
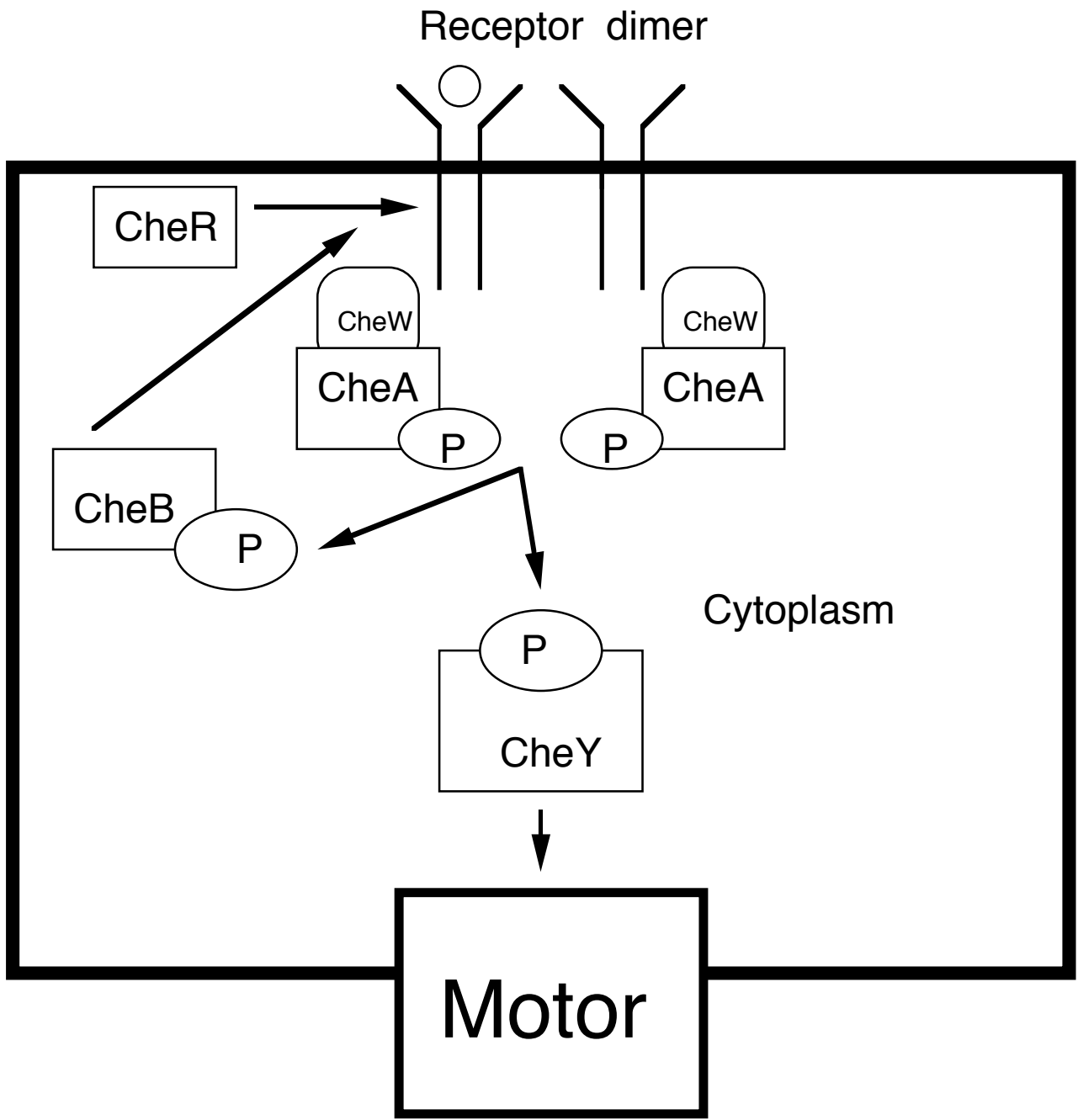
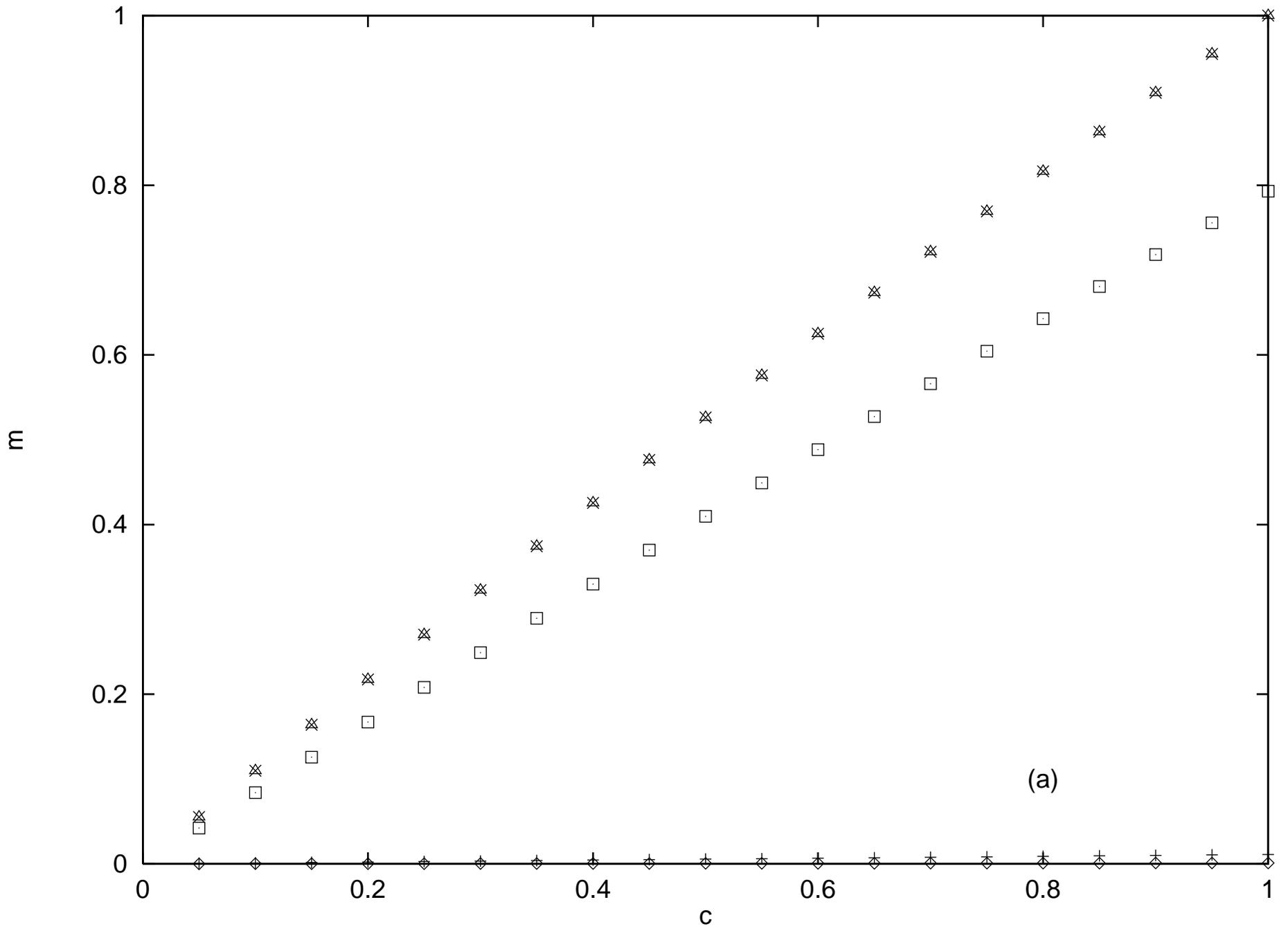
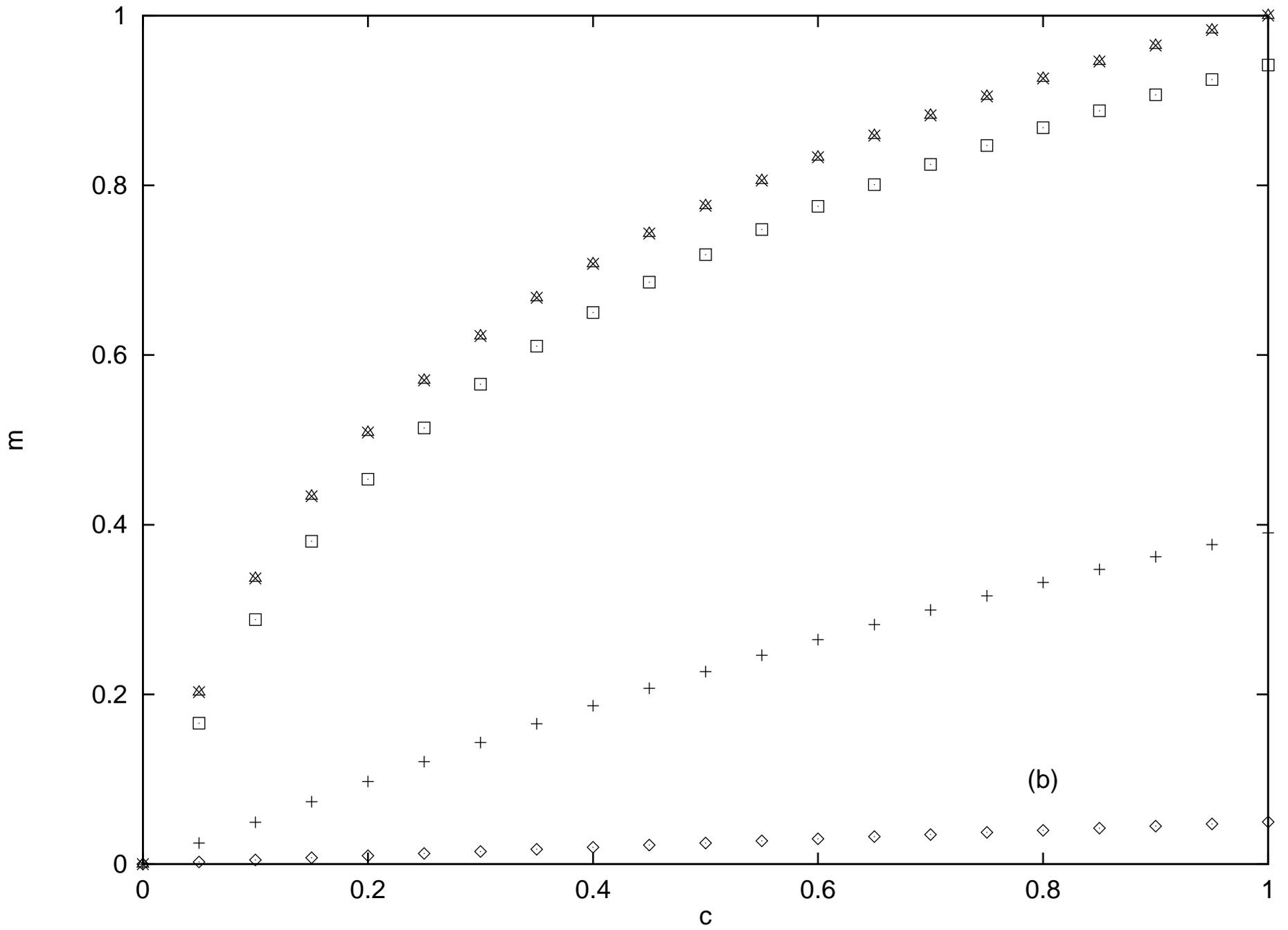
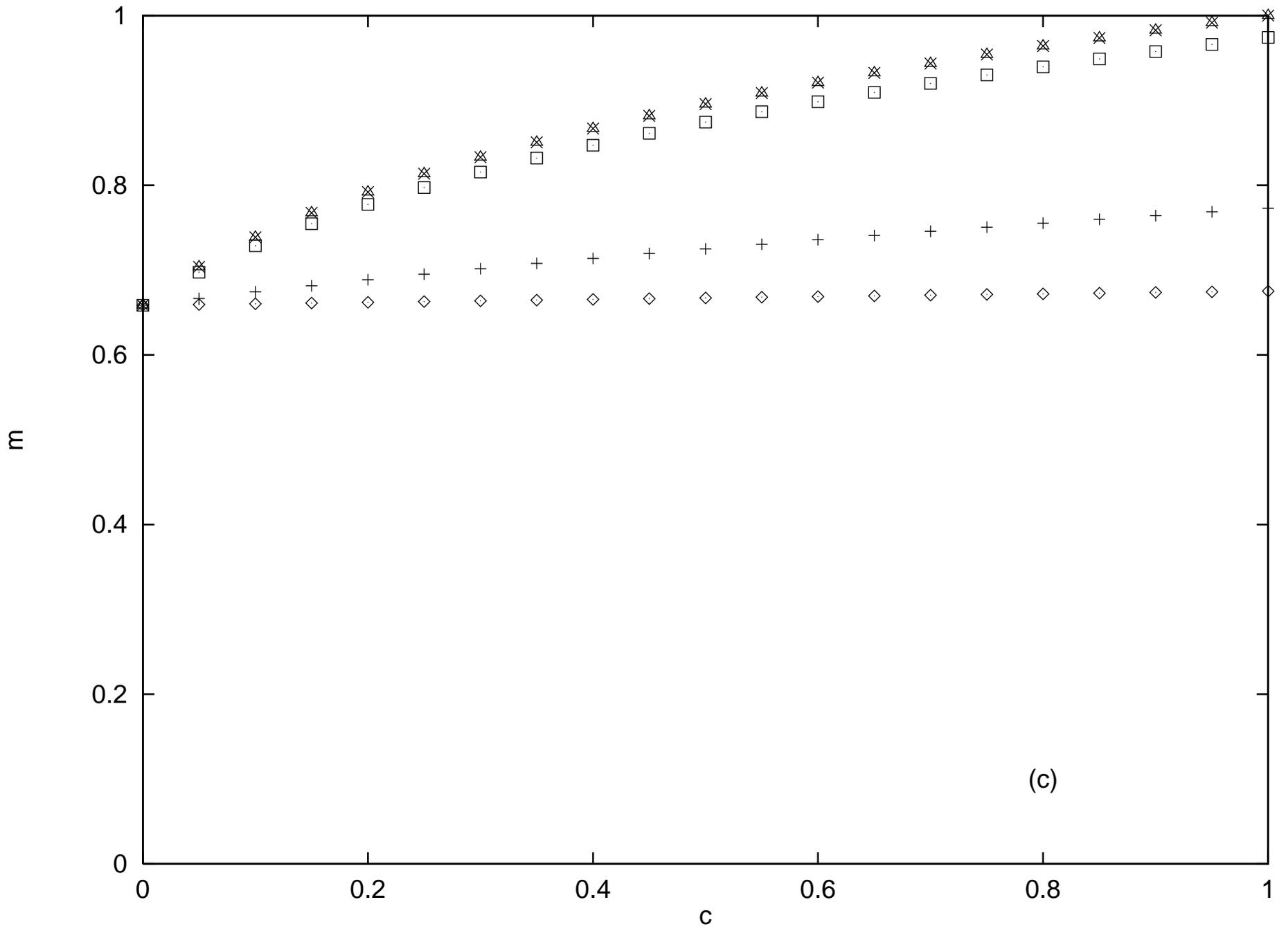
\end{document}